\documentclass[prd, twocolumn, showpacs, superscriptaddress, floatfix, nofootinbib]{revtex4}

\usepackage[pdftex]{graphicx}

\usepackage{epsf}
\usepackage{amsmath}
\usepackage{amssymb}

\usepackage{graphicx}% Include figure files
\usepackage{dcolumn}% Align table columns on decimal point
\usepackage{bm}% bold math
\def\be{\begin{equation}}
\def\ee{\end{equation}}
\def\bea{\begin{eqnarray}}
\def\eea{\end{eqnarray}}

\begin{document}

\title{Probing Particle Physics from Top Down with Cosmic Strings}

\author{Robert H. Brandenberger}
\email{rhb@physics.mcgill.ca}
\affiliation{Physics Department, McGill University, 3600 University Street, Montreal, QC, H3A 2T8, Canada}


\begin{abstract}

Making use of the wealth of new observational data coming from the sky it is
possible to constrain particle physics theories beyond the Standard Model. One
way to do this is illustrated in this article: a subset of theories admits cosmic
string solutions, topologically stable matter field configurations. In these
models, a network of cosmic strings inevitably forms in the early universe
and persists to the present time. The gravitational effects of these strings
leads to cosmological signatures which could be visible in current and future
data. The magnitude of these signatures increases as the energy scale of the
new physics involved in cosmic string formation increases. Thus, searching
for cosmological signatures of strings is a way to probe particle physics
model ``from top down", as opposed to ``from bottom up" as is done using
data from accelerators such as the Large Hadron Collider. Different ways
of searching for cosmic strings are illustrated in this article. They include
Cosmic Microwave Background temperature and polarization anisotropy maps, 
Large Scale Structure optical and infrared surveys, and 21cm intensity maps.

\end{abstract}

\maketitle

\newcommand{\eq}[2]{\begin{equation}\label{#1}{#2}\end{equation}}

\section{Introduction}
\label{Introduction}

Observational cosmology is in its golden age. Improved instrumentation
has allowed observers to explore the universe to much greater depth
than ever before. Optical and infrared telescopes yield maps of the distribution
of galaxies to larger distances, microwave experiments have provided
accurate cosmic microwave anisotropy (CMB) temperature maps, and
promise to give us CMB polarization maps in the near future.  There are
also prototype 21cm telescopes which promise to yield three-dimensional
intensity maps of the distribution of neutral hydrogen. This data is
giving us precise information about the inhomogeneities in the universe.

As has been realized for some time, the seeds for the current structure
of the universe had to have been laid down in the very early universe.
Inflationary cosmology \cite{Guth} is one paradigm for generating structure
in the universe, and alternatives such as string gas cosmology \cite{SGC},
the Ekpyrotic scenario \cite{Ekp} or the ``matter bounce" scenario \cite{Fabio}
exist. These scenarios all involve physics at energy scales much larger
than those currently explored in terrestrial accelerator experiments, scales
which in fact could approach the Planck and superstring scales. Thus, it
is natural to ask whether cosmological data can be used to explore particle
physics beyond the Standard Model. 

This article will survey one way of probing the connection between particle
physics and cosmology. It is based on the fact that in a large class of
particle physics theories beyond the Standard Model there are topologically
stable field configurations called {\it cosmic strings} which are predicted
to form in the early universe and which survive until the present time \cite{Kibble}.
They correspond to lines in space with trapped energy density. The gravitational
effects of this energy leads to distinct cosmological signatures. Searching
for these signatures is thus a way to probe particle physics beyond the
Standard Model. The magnitude of the signals is proportional to the tension 
of the strings, which in turn is proportional to the square of the energy scale 
characteristic of the string. Hence, looking for cosmic strings in the sky
is a way to probe particle physics ``from top down", i.e. with the tightest
constraints on high energy physics processes. Searching for cosmic strings
in the sky is thus an approach to test particle physics which is complementary to
terrestrial accelerator experiments which test models ``from bottom up",
i.e. where the tightest constraints are on low energy scales (see \cite{RHBCSrev}
for a more technical recent review of this topic). 

In the following section, the basics of cosmic strings will be reviewed.
Section 3 introduces the main effects by which strings lead to
cosmological signatures. Sections 4 - 7 then focus on specific observational
windows: large-scale structure, CMB temperature and polarization maps,
and 21cm redshift surveys.

A word on notation: units in which the speed of light, Boltzmann's
constant and Planck's constant are set to 1 will be used. The metric of
the background homogeneous and isotropic space-time is
\be
ds^2 = dt^2 - a(t)^2 d{\bf x}^2 \, ,
\ee
where $t$ is cosmic time, ${\bf x}$ are comoving spatial coordinates
(coordinates painted onto the expanding space, coordinates which are
constant for particles at rest in the expanding universe), and $a(t)$ is
the scale factor which describes the expansion of space. A crucial
length scale when considering the formation of structure in the universe
is the {\it Hubble radius}, the inverse expansion rate. On sub-Hubble
scales, fluctuations typically oscillate like they do in Minkowski
space-time, on super-Hubble scales they are frozen in and squeezed
(see e.g. \cite{MFB} for a comprehensive review of the theory
of cosmological perturbations and \cite{RHBfluctrev} for an introductory
overview). As a final point, we remark that it is common in cosmology
to label time $t$ by cosmological redshift $z(t)$, a measure of the
expansion of space between time $t$ and the present time. The relation
is
\be
z(t) + 1 = \frac{a(t_0)}{a(t)} \, ,
\ee
where $t_0$ is the present time.
 
\section{Cosmic String Basics}

According to quantum field theory, our best description of matter at
high energies, matter is described in terms of fields which live in space
and evolve in time. A particle corresponds to a localized point-like fluctuation
of the corresponding field, in the same way that a phonon is a localized
point-like excitation in a crystal. Some crystals admit the possibility of
stable defect lines, lines at which the orientation of the atoms in the crystal
changes abruptly. In a similar way, certain quantum field theories allow
the existence of stable field configurations with linear defects. Such defect
lines are called {\it cosmic strings}. In the same way that not all crystalline
materials admit defect lines, not all field theories admit cosmic strings.
For example, the Standard Model of particle physics does not admit
string configurations which are stable in the vacuum. However, many
extensions of the Standard Model do admit such defects.

In the same way that in condensed matter systems which admit defect
solutions, such solutions inevitably form during a fast crystallization
process, in a quantum field theory model which admits cosmic string
solutions, a network of such defect lines inevitably forms during the
cooling process in the early universe. This is the famous
Kibble mechanism \cite{Kibble2}.

In the simplest models which yield cosmic strings, the field configuration
can be visualized as a vector of fixed length $\eta$
which can point in any direction in a two-dimensional plane. For
example, consider a complex scalar field $\phi$ with potential energy
given by
\be
V(\phi) = \lambda \bigl( |\phi|^2 - \eta^2 \bigr)^2 \, ,
\ee
where $\eta$ is the characteristic energy scale of the field
theory and $\lambda$ is a positive number (the so-called ``self-coupling
constant").  In the high temperature plasma of the very
early universe, the scalar field will have sufficient thermal energy
to cross the potential energy barrier at $\phi = 0$. However,
when the temperature falls below a critical temperature
$T_c$ which is given by the energy scale $\eta$, the field
no longer has enough energy. We say that at the temperature
$T_c$ a ``phase transition" takes place. Below $T_c$.
then in order to minimize the energy, the 
field would then like to be in a ground state with $|\phi| = \eta$. However, by
causality the orientation of the $\phi$ vector in the complex
plane must be random on scales beyond which no information
has had time to propagate since the time of the Big Bang. Hence, on such
scales there is a probability of order one that if one considers a
circle ${\cal C}$ in space on that scale, then the field vector $\phi$ will
circle the point $\phi = 0$. By continuity, it then follows that
there needs to be at least one point on any disk bounded by ${\cal C}$
with $\phi = 0$. These points connect to a line, and this is the
center of the cosmic string. The energy density $\mu$ of the string is obtained
by minimizing the sum of potential and gradient energies and is given 
by 
\be
\mu \simeq \eta^2 \, ,
\ee
where the dependence on the coupling constant $\lambda$ cancels out.
As the observant reader will have noticed, the above argument
is based on causality and holds at all temperatures below
the critical temperature, and thus at all times up to the present
time. Thus, and this is the second aspect of the Kibble mechanism,
in particle physics models which admit cosmic string solutions, a
network of strings will persist until the present time.

The detailed form of the distribution of the cosmic string network
must be studied by means of numerical simulations (see e.g.
\cite{CSsimuls} for a selection of references on such simulations).
However, the qualitative aspects can be derived using analytical
considerations (see e.g. \cite{VilShell, HK, RHBrev} for early
reviews on the cosmology of cosmic strings). 
First of all, note that cosmic strings cannot have
ends. Hence, they are either part of an infinite string network, or
else there are loops. The infinite string network is described by
a correlation length $\xi(t)$ which depends on cosmic time $t$
and gives the mean separation and mean curvature radius
of the long strings (in the ``one scale" model of the string network,
these two lengths are the same). Due to the tension, a curved
string will typically have velocity in direction perpendicular to
the string whose magnitude $v$ is a significant fraction of the
speed of light. As a consequence of the induced motion,
there will be frequent string intersections which both produce
string loops and lead to an increase in $\xi(t)$. On the other
hand, by causality $\xi(t)$ is bounded from above by the
horizon distance $t$ (we are using units in which the speed of
light is set equal to 1). The dynamics can be studied by means
of a Boltzmann equation (see e.g. \cite{RHBrev}) which in
fact yields 
\be
\xi(t) \sim t \, ,
\ee
where the proportionality constant must be determined numerically.
It is found that a horizon volume has on the average $N$ long
strings passing through it, where $1 < N < 10$ based on simulations
by different groups.

To summarize, it follows from analytical arguments that the
cosmic string network will at times much later than the time
of the phase transition take on a ``scaling solution" according
to which the statistical properties of the string network are
time-independent if all lengths are scaled to the horizon distance
$t$ (see Figure 1 for an illustration of the scaling solution).

\begin{figure}
\includegraphics[height=12cm]{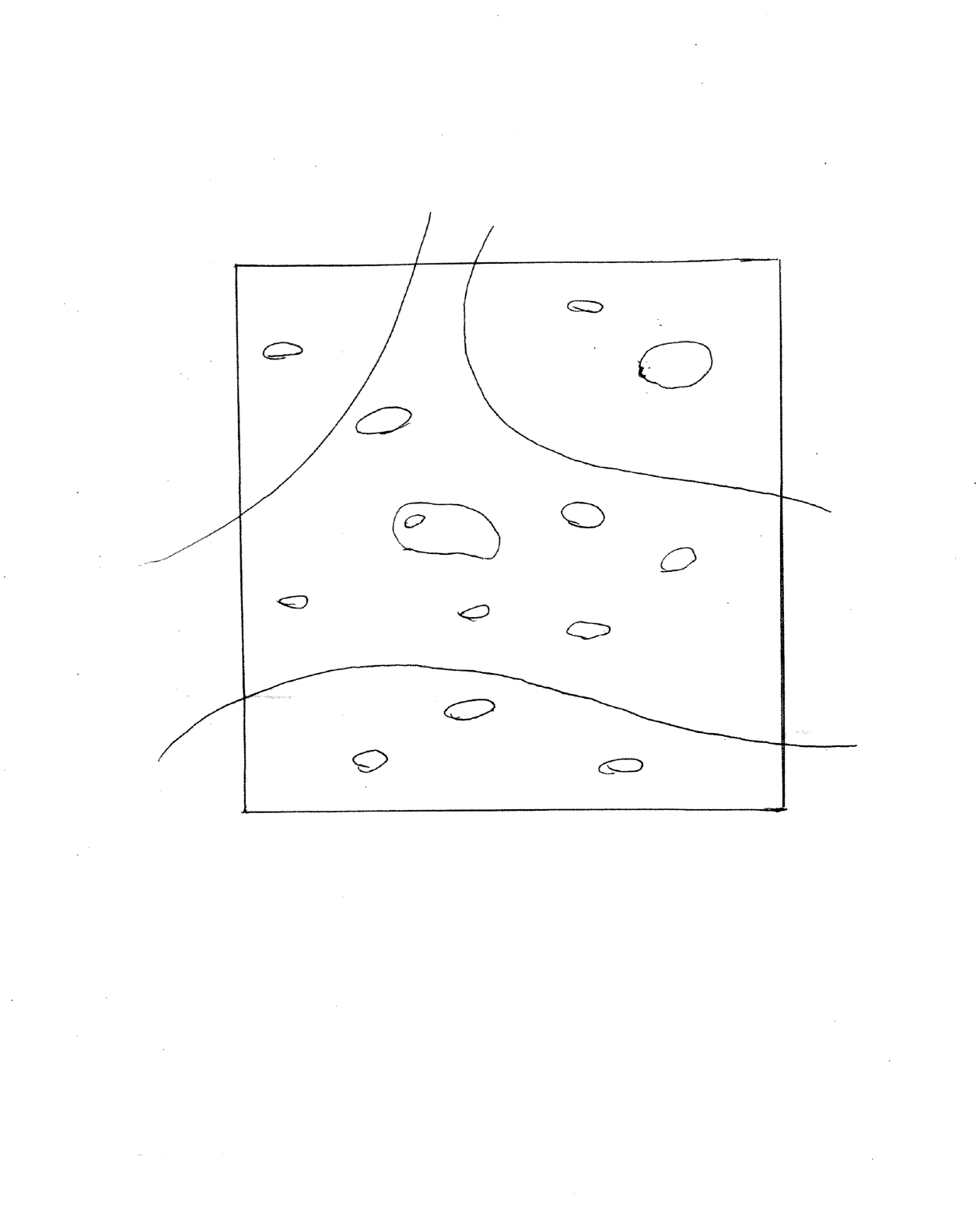}
\caption{Sketch of the cosmic string scaling solution. What is shown
is the distribution of strings in a fixed Hubble volume (projected onto a
plane).}
\label{fig1}
\end{figure}

The cosmic string loops surviving at late times are the remnants
of the interactions between long string segments. Their 
distribution also scales, and is characterized by a maximal
radius $\alpha t$, where $\alpha$ is a positive number of the
order but somewhat smaller than 1. As a function of
radius $R$, the number density $n(R, t)$ of loops scales as $R^{-5/2}$
for loops formed before the time of equal matter and radiation
(see e.g. \cite{TB}).
String loops oscillate and emit gravitational radiation \cite{CSgravrad}. This
effect leads to the fact that the number density $n(R, t)$
becomes constant below a cutoff radius given by
\be
R_c = \gamma G \mu t \, ,
\ee
where $\mu$ is the string tension, $G$ is Newton's gravitational
constant and $\gamma$ is a real number whose value is
determined in numerical simulations to be about $100$.

Since a cosmic string arises in a relativistic quantum field
theory, its mass per unit length is equal in magnitude to
the tension. The gravitational effects of a string are hence
proportional to $\mu$. They are typically parametrized by
the dimensionless constant $G \mu$ whose value is in
the range $10^{-8} - 10^{-6}$ for cosmic strings formed
in ``Grand Unified" models of particle physics, and which
are predicted to be in the range $10^{-12} - 10^{-6}$ for
cosmic superstrings \cite{Witten}, fundamental strings
which can be stabilized as macroscopic objects in certain
string models \cite{CMP}.

Since in theories which admit cosmic strings a network of
such strings survives until the present time, these strings
will have observational consequences which can be searched
for in observations (this was first realized in \cite{Zel, Vil, Kibble}).
This will be the topic of the rest of this review.

\section{Main Effects}

The network of cosmic strings at late times contains string
loops and network of infinite strings with a correlation length
$\xi(t)$ - we will call the latter the ``long strings". Both
loops and long strings lead to gravitational effects.
Since, according to the current cosmic string evolution
simulations most of the energy of the string network
is in the string loops, the loops will dominate the
cosmological effects. However, the long strings lead
to more characteristic signatures and hence might
be easier to detect in the sky. An important message
of the following sections is that it is better to look for
the signal of strings in position space rather than in
Fourier space. The first reason is that in Fourier space the 
specific signals of the strings are washed out. The second
reason is that (Fourier space) power
spectra typically depend quadratically on the
constant $N$ which parametrizes our ignorance of the
exact nature of the scaling solution, whereas the position
space signals in the idealized case are independent on $N$.

\subsection{Effects of string loops}

A string loop will oscillate and slowly emit gravitational
radiation. Viewed from a distance larger than the loop radius,
the time-averaged gravitational effect of the loop is like that
of a point mass of magnitude
\be
M(R) = \beta \mu R \, ,
\ee
where the deviation of the number $\beta$ from $2 \pi$
describes the difference of the loop shape from being circular.
Hence, cosmic string loops will be the seeds of roughly
spherical accretion (for simplicity we here neglect
the initial center of mass motion of the loops). 
Before the time $t_{eq}$ of equal
matter and radiation, the accretion of baryons and dark
matter is suppressed. Thus, the main growth starts at $t_{eq}$.

As long as there is enough matter to feed the growth, the
growth about string loops will proceed independently.
Loops of different radii $R$ will lead to objects of different
final masses $M_f(R)$
\be
M_f(R) \simeq \beta \mu R \frac{z_{eq}}{z_{to}} \, ,
\ee
where the last factor describes the gravitational accretion
from time $t_{eq}$ (with corresponding redshift $z_{eq}$)
until the redshift $z_{to}$ when accretion stops, e.g.
the present time for small values of $G \mu$ and for
loops sufficiently displaced from any larger loop.

The initial work on cosmic strings and structure formation \cite{TB}
mostly focused on the effects of string loops, and 
made the hypothesis that cosmic strings were responsible
for all of the inhomogeneities and anisotropies in the universe,
and thus would be an alternative to cosmic inflation as an explanation
for the origin of structure. As a consequence of the scaling
distribution of cosmic strings, the spectrum of density fluctuations
from strings is scale-invariant \cite{BT}. For a value of
$G \mu \simeq 10^{-6}$ the spectrum would have the right
overall amplitude to explain the large-scale structure (LSS) data. 

However, the fluctuations
start out as entropy fluctuations on super-Hubble scales, rather
than adiabatic perturbations in the case of inflation. They are
incoherent and active (i.e. the induced adiabatic component
is growing on super-Hubble scales). As a consequence of
these facts, the CMB anisotropies on small angular
scales are predicted to be different from the results for primordial
adiabatic fluctuations: there are no acoustic oscillations in the
CMB angular power spectrum \cite{incoherent}, but only one
broad Doppler peak. 

Once the acoustic oscillations in the CMB angular power spectrum
were discovered by the Boomerang \cite{Boomerang} experiment 
and later confirmed by data from the WMAP satellite mission \cite{WMAP}
it became clear that cosmic strings could not account for all of
the structure in the universe. This leads to an upper bound on the
cosmic string tension. The most robust current bound on the string tension
comes from the recent CMB angular power spectrum measurements,
in particular the South Pole Telescope (SPT) \cite{SPT}, the Atacama
Cosmology Telescope (ACT) \cite{ACT} and the Planck mission \cite{Planck}.
The bound on the string tension from a combination of these
experiments is \cite{CSbound} (see \cite{previous} for earlier limits)
\be
G \mu < 1.5 \times 10^{-7} \, ,
\ee
where the coefficient in fact depends on the constant $N$ describing
the string scaling solution (which has been mentioned earlier). 
This implies that cosmic strings can be responsible for at most 
about $5\%$ of the power of density fluctuations in
the current universe. The dominant source of inhomogeneities must be
a different process such as inflation \cite{MukhChib} or one of its alternatives
(see e.g. \cite{RHBalt} for a recent comparison between inflation and some of
its alternatives).

From the particle physics point of view, however, there are good reasons to
expect cosmic strings to be present, even if the universe underwent a period
of inflation. In many supergravity models of inflation, strings are produced at
the end of inflation \cite{Rachel}, and in brane inflation models motivated
by superstring theory, similarly a network of strings is the result of the termination
of inflation \cite{Tye}. In string gas cosmology, a network of cosmic
superstrings may survive. Hence, looking for cosmological signals of strings
provides a good way to probe particle physics beyond the Standard Model.

Cosmic string loops will  accrete matter in a roughly spherical
manner. String loops could, for example, contribute to the formation of
ultra-compact mini-halos embedded within a galaxy halo \cite{Berez}.
For a more general recent discussion of the role of cosmic string loops
see e.g. \cite{Shlaer}. However, in this article we will focus on signatures of 
long strings since their geometric patterns are more distinctive.

\subsection{Kaiser-Stebbins effect}

Because of the fact that a cosmic string has relativistic tension (tension equal
in magnitude to the energy density), the gravitational effects of long straight
strings are very special. There is in fact no local gravitational force exerted by the
string. On the other hand, globally space is non-trivial - space perpendicular
to a cosmic string is conical with a ``deficit angle" $\alpha$ of magnitude
proportional to the tension \cite{deficit}
\be 
\alpha = 8 \pi G \mu \, .
\ee

The conical structure of space perpendicular to a long string leads
to lensing of light passing on different sides of the string. This,
in turn, gives rise to a specific signature of long straight strings
in CMB temperature maps \cite{KS, Gott}: consider a source of
light behind a string moving with velocity $v_s$ in the plane
perpendicular to the string, from the point of view of an observer at rest 
in the frame of the cosmological background. Light from this source
will reach the observer on two different paths which pass
on different sides of a string. There will be a relative Doppler
shift in the frequency which the observer sees. Specifically,
this effect applies to the CMB background radiation. Hence,
a string will produce a line in the sky across which there is
a relative temperature jump of magnitude
\be \label{jump}
{{\delta T} \over T} \, = \, 8 \pi \gamma(v_s) v_s G \mu \, ,
\ee
where $\gamma_s$ is the relativistic gamma factor associated
with the velocity $v_s$. For an illustration of this effect see
Figure 2.

\begin{figure}
\includegraphics[height=12cm]{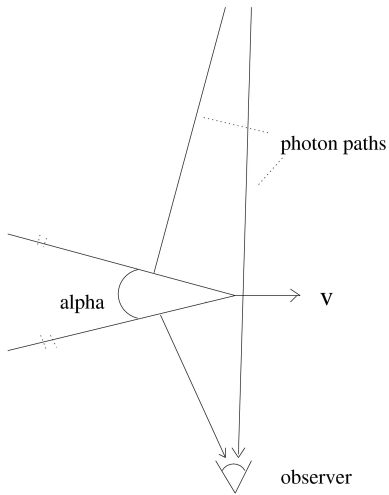}
\caption{Geometry of space perpendicular to a
long straight string segment. Flattened out, there
is a deficit angle whose magnitude is proportional
to $G \mu$. A string moving with
velocity $v_s$ in transverse direction leads to a relative
Doppler shift between photons passing on different sides of the string.}
\label{fig2}
\end{figure}

For cosmic strings formed during a phase transition in the early
universe, the conical structure of space extends only a finite distance
away from the string. As derived in \cite{Joao}, this distance is
the time that light can travel between string formation and the
time $t$ which is being considered. Beyond that distance, space
perpendicular to the string rapidly converges to being Euclidean.
Hence, a string segment leads to a temperature surplus on one
side of the string and a deficit on the other extending a distance
corresponding to the comoving horizon at the time the light
is passing by the string.

\subsection{Cosmic string wakes}

The conical structure of space perpendicular to a long straight string
is also responsible for the second important effect, the ``cosmic string
wake". We consider the same string segment moving with velocity
$v_s$ in direction perpendicular to its tangent vector. It is moving
through the cosmic gas. From the point of view of an observer behind
the string, a velocity perturbation of the gas towards the "world
plane" of the moving string (the plane spanned by the tangent vector
to the string and the velocity vector $v_s$) is induced whose
magnitude is
\be
\delta v = 4 \pi G \mu v_s \, .
\ee
Hence, a wedge-shape region of overdensity 2 behind the string is
induced, the ``wake" \cite{Silk}. The wedge is a three-dimensional object.
A string segment of length $c_1 t$, where $c_1$ is a constant,
induces a wake of dimensions
\be
c_1 t \times v_s \gamma_s t \times 4 \pi G \mu \gamma_s v_s t \, ,
\ee
where the second dimension is the depth and the third is the mean
width, the width being zero at the point where the string is located,
and twice the above value at the far end (see Fig. 3).

\begin{figure}
\includegraphics[height=5cm]{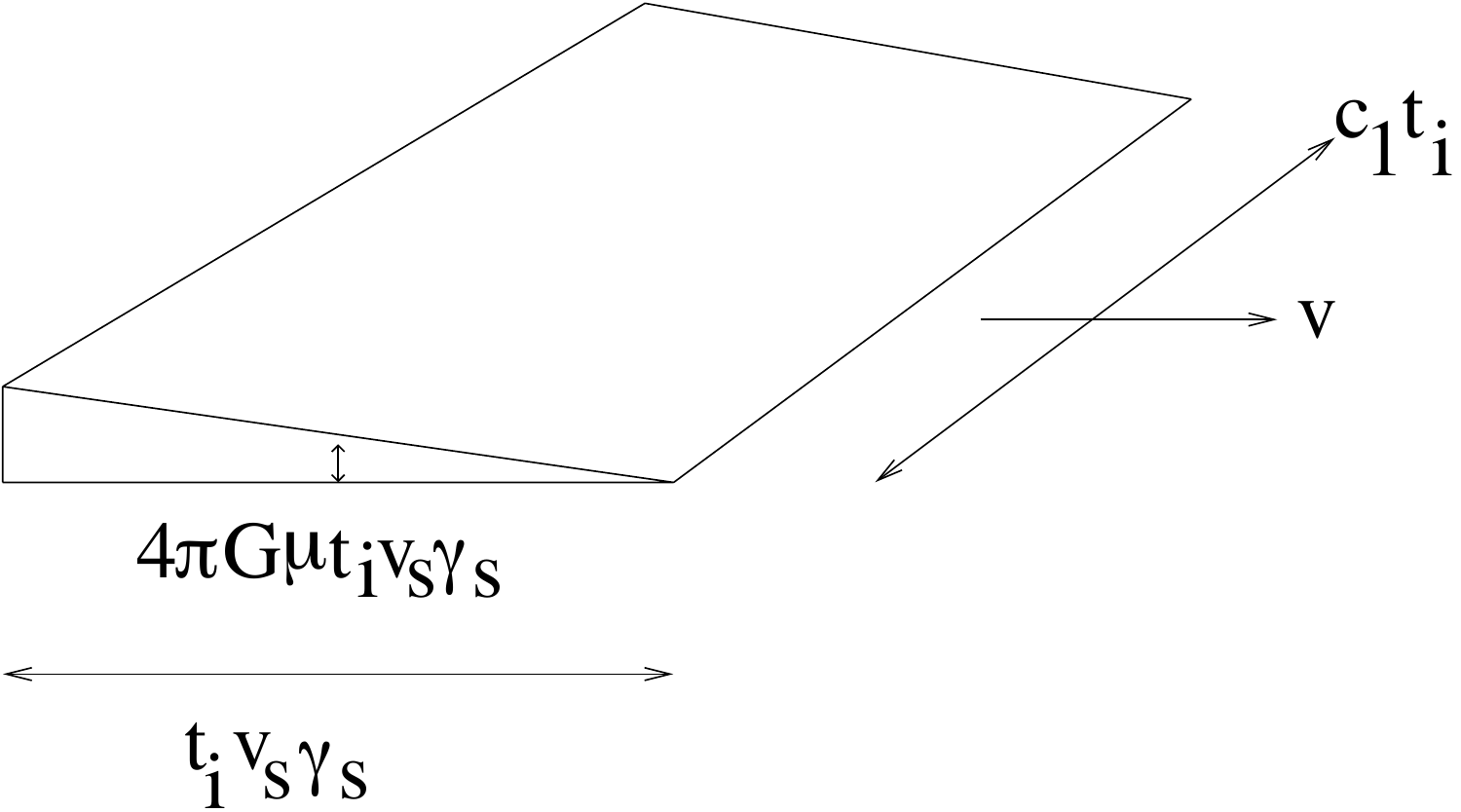}
\caption{The geometry of a cosmic string wake.}
\label{fig3}
\end{figure}

Once formed, a wake will grow in thickness by gravitational
accretion, and it will expand in the planar directions following the
expansion of space. A long string present at time $t_i$ (taken to
be larger than $t_{eq}$) will
create a wake whose physical dimensions at a later time $t$ will
be
\be \label{size}
c_t t_i \frac{a(t)}{a(t_i)} \times v_s \gamma_s t \frac{a(t)}{a(t_i)} \times
4 \pi G \mu \gamma_s v_s t_i [\frac{a(t)}{a(t_i)}]^2 \, ,
\ee
where the extra factor of $\frac{a(t)}{a(t_i)}$ in the third term comes
from the growth of the width via gravitational accretion. The
magnitude of this factor is easy to understand: it represents the
linear cosmological perturbation theory growth of fluctuations.
An improved derivation is based on the Zel'dovich approximation
\cite{Zeld}. The thickness is defined as the physical radius of
the mass shell which has just ``turned around" and is beginning
to collapse onto the central plane of the wake (initially every
mass shell of particles above the wake is moving away from
the wake because of cosmological expansion. The Zel'dovich
approximation has been applied to cosmic string wakes in
\cite{Zelwakes}.

A key point is that cosmic string wakes correspond to nonlinear
structures present already at the earliest times. This is in contrast
to what happens in theories with Gaussian fluctuations such as
those produced by inflation. In those theories fluctuations remain
in the linear regime until times close to the epoch of reionization
at redshifts of about 10. Hence, observational windows which probe
nonlinear structures at high redshifts are more likely to find signatures
of cosmic strings than surveys which probe cosmological structures
at the present time. 

\section{Cosmic Strings and Large-Scale Structure}

In this section we will discuss signatures of cosmic strings in
the large-scale structure observed through optical and infrared
telescopes. We will focus on the role of the infinite string network
since the long strings give rise to more distinctive geometrical
patterns.

The long strings form a network of infinite extent. However,
for analytical studies it is convenient to break up this network
into a set of segments, each of length $\xi(t)$. As the
infinite strings self-intersect and split off loops, the set of
string segments changes. The time scale for string intersections
is the Hubble expansion time scale $t$. Hence, string segments
can be viewed as statistically independent on time scales
larger than $t$. This idea gives rise to a toy model \cite{Periv}
for the distribution of long strings: we divide the time period
under consideration (usually starting either at $t_{eq}$ or else
at the time $t_{rec}$ of recombination and ending at the present
time $t_0$) into Hubble time steps. In each time step (with
initial time $t_i$, where $i$ is the index of the time step), we take
a fixed number $N$ of straight string segments of length $\xi(t)$,
and moving with velocity $v_s$. The centers, orientations, and
velocity unit vectors are chosen at random.

Each string segment present at time $t_i$ will produce a wake
whose dimensions at a later time $t$ are given by (\ref{size}).
The wakes will thus imprint a planar topology to the distribution
of dark matter in the universe. In the ``old days" of cosmic string
cosmology the effects of wakes on the large-scale structure of
visible matter was analyzed in many works, e.g. in \cite{wake}.
Since there is only logarithmic growth of the density fluctuations
before $t_{eq}$, the time interval to consider for studies of large-scale
structure is the interval between $t_{eq}$ and the present time.
As a consequence of the scaling distribution of the string network,
the overall power spectrum of induced density fluctuations is
scale-invariant \cite{TB}. The specific signatures of cosmic strings
cannot be seen in the two point correlation function, but only
in non-Gaussian statistics. A better approach than naively computing
three- and four-point correlation functions is to look at topological
statistics such as Minkowski functionals \cite{Mink}. This has been
done in some early studies in \cite{Mitsouras}.

For the small values of $G \mu$ which are now allowed by the
bounds on the cosmic string tension, it was recently shown that
halos which are induced by the accretion of matter onto wakes
are too small and have too low a temperature to induce star formation.
Hence, they will be difficult to detect directly \cite{Duplessis}, in
spite of the fact that at higher redshifts the power spectrum
of nonlinear matter becomes dominated by the string wakes.
A more detailed study of this issue is in preparation \cite{Yuuki}.

\section{CMB Temperature Anisotropies from Strings}

The cosmic microwave background (CMB) provides an image
of the distribution of matter at the time $t_{rec}$ of recombination,
a time when the fluctuations are in the linear regime and
precise analytical calculations are possible. As already
mentioned in Section 3, long string segments lead to a clear
geometrical signature in CMB temperature anisotropy maps:
a line in space across which the temperature jumps by
(\ref{jump}). 

In fact, due to the finite depth of the conical
geometry about the string, a string segment whose
world sheet is crossed by our past light cone at
time $t_i$ leads to a rectangular
pattern in the sky with $\delta T \neq 0$. In the direction
on the sky corresponding to the string motion, there is
a rectangle of depth 
\be
\frac{1}{2} \theta_i v_s \gamma_s
\ee
in front of the string with a positive temperature fluctuations
$\delta T$ (given by (\ref{jump}), whereas behind the string
there is a rectangle of similar size the a negative temperature
fluctuation of equal magnitude. The width of these rectangles
in direction tangent to the string is given by
\be
c_1 \theta_i \, ,
\ee
where $\theta_i$ is the angle corresponding to the comoving
Huubble radius at the time $t_i$, and is given by
\be
\theta_i = 90^{o} [z(t_i) + 1]^{-1/2} \, .
\ee
A sketch of the pattern of an individual string segment is given in
Figure 4.

\begin{figure}
\includegraphics[height=12cm]{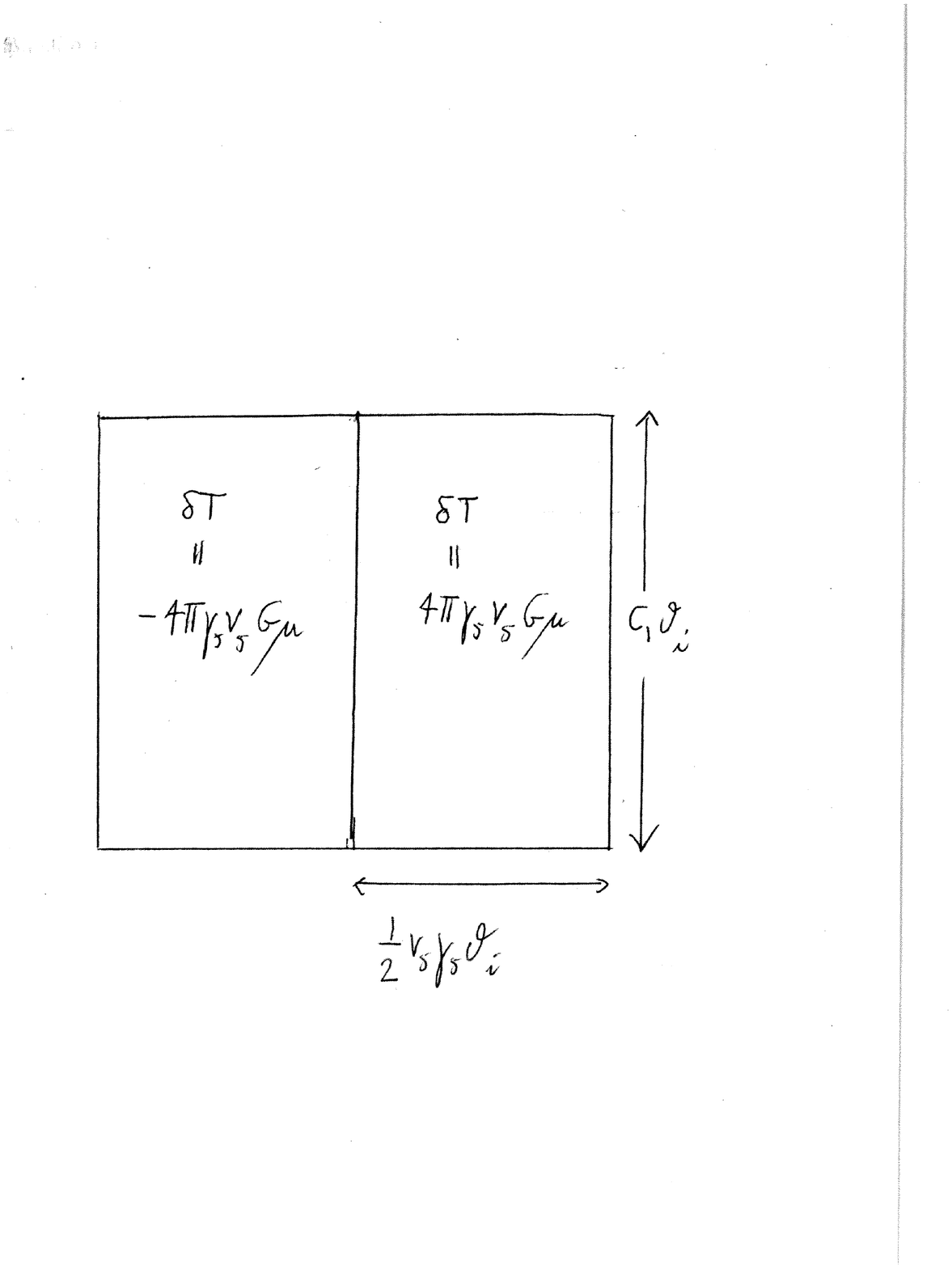}
\caption{The temperature pattern on the sky induced by a single
string segment.}
\label{fig4}
\end{figure}

A scaling network of string segments produces a CMB temperature
anisotropy map which can be constructed as follows \cite{Danos}: given
a patch of the sky which is being modelled, we compute the number of
string segments present at each Hubble expansion time $t_i$ which
are intersected by the portion of the past light cone which the patch
of the sky corresponds to. We then choose random centers, directions
and velocity vectors for each segment and construct the resulting
temperature rectangle in the sky given by the above-mentioned
dimensions. The CMB map is a superposition of the rectangles
given by the individual string segment. Figure 5 (taken from \cite{Danos})
presents an image corresponding to a $10^{o} \times 10^{o}$ patch of the sky. 
Note the large number of small rectangles (early $t_i$) compared to the 
larger ones (late $t_i$).

\begin{figure}
\includegraphics[height=12cm]{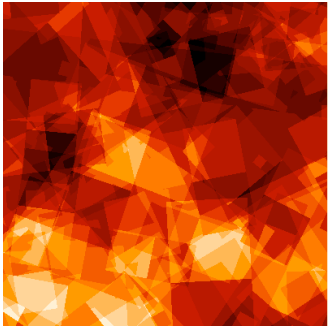}
\caption{CMB anisotropy map for a $10^{o} \times 10^{o}$ 
patch of the sky at $1.5^{'}$ resolution (the specifications 
are chosen with applications to the SPT and ACT 
telescope data in mind) in a model in which the fluctuations are 
given by a scaling distribution of cosmic strings. The
color coding indicates the amplitude of the temperature
anisotropy.} 
\label{fig5}
\end{figure}

The best current limits on the cosmic string tension come from
comparison of the angular power spectrum of CMB temperature
anisotropies \cite{CSbound}. By making use of topological statistics such
as Minkowski functionals or edge detection algorithms, improved
limits should be obtainable. Initial studiesof CMB maps using Minkowski 
functionals have been done in \cite{MinkCMB}, and an application of the 
Canny edge detection algorithm \cite{Canny} to constructed maps
including the contributions from string wakes has shown \cite{Danos, Berger}
that an improvement of the limit by one order of magnitude might
be achievable with current SPT data.

\section{CMB Polarization Patterns from a Cosmic String Wake}

CMB radiation can be polarized. The polarization data of CMB
anisotropies carries interesting cosmological information. 
Polarization can be decomposed into two independent components
which are called E-mode and B-mode polarization. The primordial 
E-mode polarization has already been 
discovered \cite{Emode}, but only upper bounds are available
on the B-mode. In the near future, experiments such as the SPT \cite{SPTPol},
ACT \cite{ACTPol} and Planck will provide us with greatly improved polarization
maps. In this context it is interesting that cosmic strings lead to a
specific pattern of CMB polarization, in particular B-mode polarization.
 
Regular adiabatic fluctuations such as those produced during
inflation do not produce any direct B-mode polarization. B-mode
polarization in such models can be generated from gravitational
waves, or from the lensing of E-mode polarization. In contrast,
cosmic strings lead to direct B-mode polarization as was
first realized in \cite{first}. The position space signal of
B-mode polarization from a string was studied in detail in
\cite{Holder1}.

CMB polarization is induced by the CMB photons from last scattering
crossing a cosmic string wake. Only the quadrupole component $Q$
of the CMB induces polarization. The wake is an overdensity of matter,
including baryonic matter. Hence, it also represents an overdensity
of free electrons. The CMB photons Compton scatter off of the free 
electrons, and this process induces CMB polarization. Each wake
which is intersected by our past light cone leads to a rectangle in
the sky with extra CMB polarization. The angular dimensions of
this wake-induced rectangle for a wake created at time $t_i$ are 
\be
[c_1 \times v_s \gamma_s ] \theta_i^2
\ee
(as long as the time $t$ when the past light cone intersects the wake
is significantly smaller than $t_0$ - for the general formula the
reader is referred in \cite{CMBPolpower}). Across the rectangle
in the sky, the polarization direction is roughly constant. The magnitude
$P$ is proportional to $Q$, to the Thompson cross section, to the width of the
wake (which in turn depends on $G \mu$ and on the times $t_i$ and
$t$), to the free electron fraction $f(t)$ at the time $t$, and to the overall
baryon energy fraction $\Omega_B$.  Inserting numbers, the result is
\cite{Holder1}
\be
\frac{P}{Q}(t, t_i) \sim f(t) G \mu v_s \gamma_s \Omega_B \bigl( \frac{z(t) + 1}{10^3} \bigr)^2
\bigl( \frac{z(t_i) + 1}{10^3} \bigr)^{1/2} 10^7 \, .
\ee
The above is the mean amplitude across the rectangle. The amplitude increases
linearly moving away from the position of the string. The pattern is
sketched in Fig. 6 (taken from \cite{Holder1}).

\begin{figure}
\includegraphics[height=7cm]{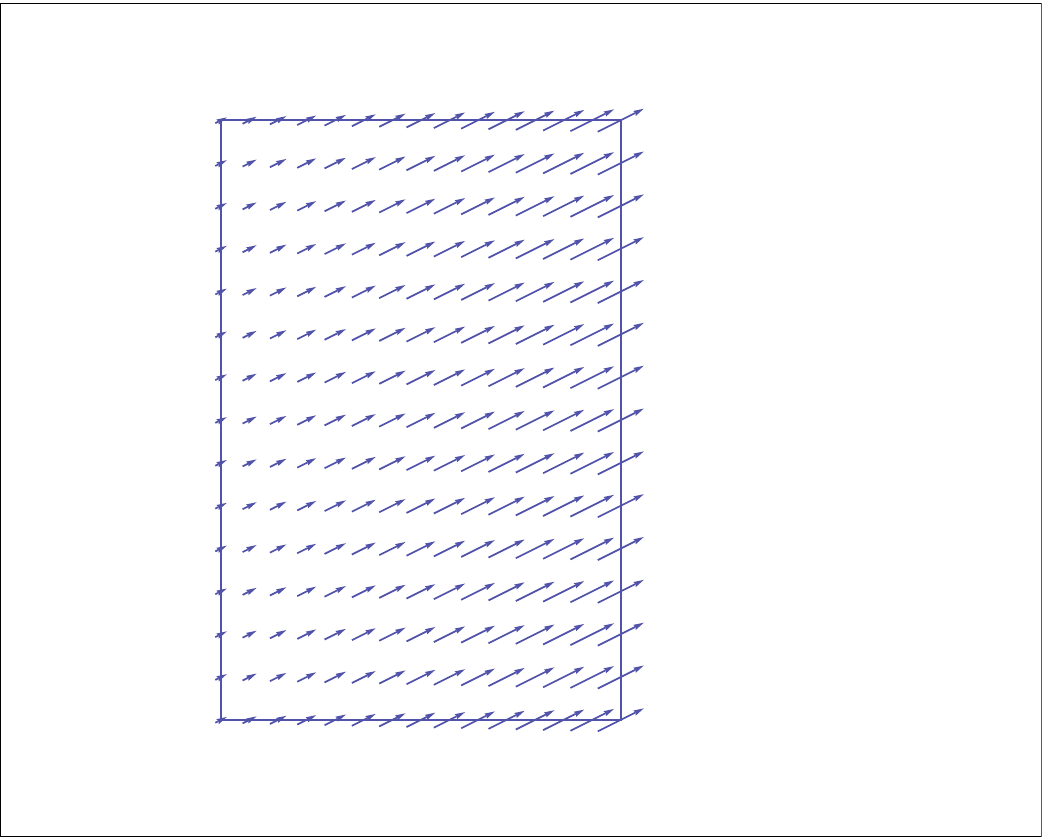}
\caption{Sketch of the polarization pattern induced by
a cosmic string wake. Shown is a portion of the sky, e.g.
$10^{o} \times 10^{o}$, and the polarization which the
wake induces. The direction of the arrows indicates
the polarization direction, the length of the arrow gives
the magnitude.} 
\label{fig6}
\end{figure}

The important point is that, as discussed in detail in \cite{CMBPolpower},
the mean amplitude of the induced E and B mode polarizations are the
same. Cosmic string wakes thus induce directly B mode polarization, unlike
standard adiabatic fluctuations. Hence, the search for B mode polarization
is a very important current goal for observational cosmology. The discovery
of such polarization could give us a lot of very important information  about
the early universe. It could confirm the existence of cosmic strings (this is
the point emphasized in this article). If any discovered B mode polarization
could be shown to be due to gravitational waves, it could shed light on
the origin of primordial gravitational waves. A roughly scale invariant
spectrum of such waves is predicted not only by inflationary cosmology, but
also by a scaling solution of cosmic string loops, and in fact with an amplitude
larger than what is predicted in the simplest inflationary models. 
A discovery of a gravitational wave spectrum with a slight blue tilt would rule 
out all inflationary models based General Relativity with matter satisfying the 
usual energy conditions, and it would \cite{BNPV} verify a key prediction of 
string gas cosmology (see \cite{holygrail} for a more detailed discussion of 
these points).

The angular power spectrum of string wake-induced B-mode CMB polarization was
computed recently in \cite{CMBPolpower}. The spectrum is 
almost degenerate with that of B-mode polarization induced by
lensing. This is another example of the message that in order to detect
signals from string wakes, one needs to analyze data in position space.

\section{Imprints of Cosmic Strings in 21cm Surveys}

The final window to probe for the possible existence of cosmic strings
which we discuss in this article is the window which 21cm redshift
surveys provide. These surveys provide, in addition to many other
purposes,  a means of probing the distribution of baryonic matter during 
the dark ages, i.e. before the time of star formation. Since 21cm
redshift maps are three-dimensional (two dimensions the angles in
the sky, the third dimension the redshift), they provide potentially much
more information than two dimensional CMB maps.

Before star formation (but after recombination), baryonic matter is almost
entirely in the form of neutral hydrogen. Neutral hydrogen has a hyperfine
transition line with a wavelength of 21cm. If photons of the cosmic
microwave background travel through a cloud of neutral hydrogen,
there will be either absorption or emission of 21cm radiation - absorption
if the CMB temperature is larger than the gas temperature in the cloud,
emission if the gas temperature is larger. Since cosmic string wakes
are overdensities of neutral hydrogen, they will lead to a larger 21cm
signal.

The 21cm signal of a single string wake was discussed in \cite{Holder2}
(see \cite{21early} for initial studies of the imprints of cosmic strings
in the 21cm background, and see \cite{Furl} for a general overview of
the physics of 21cm surveys). The geometry of the string-induced
pattern is very special: extended in the two angular directions and
thin in redshift direction. The geometry is sketched in Figure 7
(taken from \cite{Holder2}). On
the left side of the figure is a space-time sketch showing the
location of the string wake as a function of time, and the past
light cone of our observer intersecting the wake. On the right side
of the figure is the corresponding signal in redshift space. The
horizonal axes on both sides of the figure are the same. The vertical
axis is time on the left graph and redshift on the right graph. The
signal of the string wake is a thin wedge, roughly but not quite
perpendicular to the redshift axis. Within the region of the wedge
there is extra 21cm absorption (or emission if $G \mu$ were to be
large). The mean thickness of the wedge depends on $G \mu$ since
it is determined by the width of the wake. On the other hand, it
turns out the the amplitude of the signal is independent of $G \mu$.
Thus, if a telescope with excellent redshift resolution were available,
the 21cm window would be an excellent one to search for lower
tension strings.

\begin{figure}
\includegraphics[height=4.5cm]{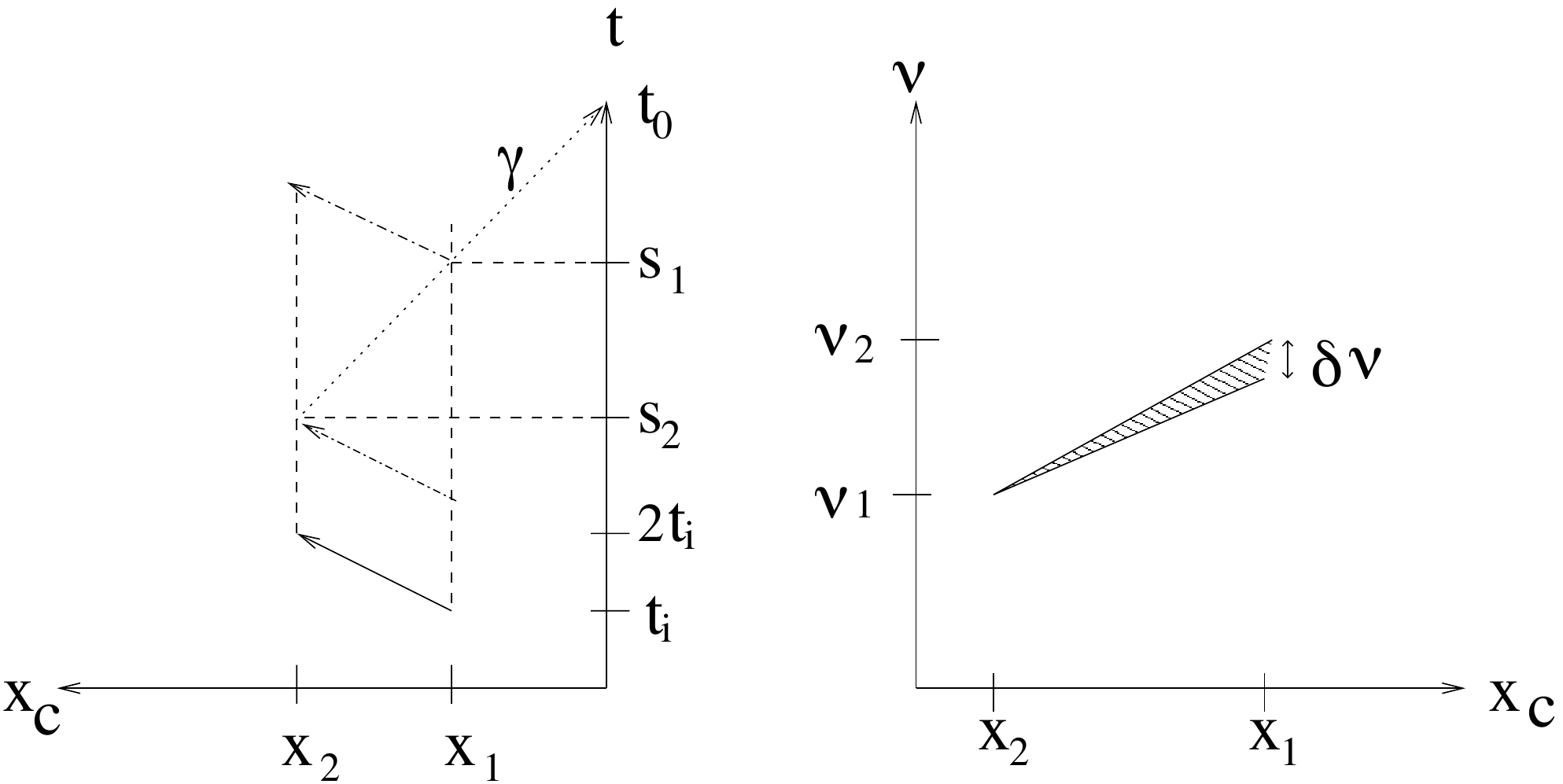}
\caption{Geometry of the 21cm signal of a cosmic string wake. The
left side shows a space-time sketch with the horizontal axis being
comoving spatial coordinate and the vertical axis conformal
time) of a  wake created at time $t_i$. The string segment ``lives'' until
the time $2 t_i$. Its initial and final positions are $x_1$ and
$x_2$, respectively. The wake extends
from $x_1$ to $x_2$. Its thickness increases linearly from 
$x_1$ to $x_2$. The
past light cone (indicated by the line labelled $\gamma$)
intersects the string wake. In the example shown, the
past light cone intersects the leading (thin) edge of the
wake earlier than the trailing edge. Thus, 21cm photons from
the leading edge are redshifted more than those from the trailing
edge. This generates a characteristic wedge of extra
21cm absorption / emission due to the string wake in
redshift maps. This is sketched on the left of the
figure. Here, the horizontal axis is the same
comoving spatial coordinate as in the left side, and
the vertical axis is the detected 21cm frequency.} 
\label{fig6}
\end{figure}

The amplitude of the effect of the string wake 
is given by
\be \label{basic1}
T_b(\nu) \, = \, 
T_S \bigl( 1 - e^{- \tau_{\nu}} \bigr) + T_{\gamma}(\nu) e^{- \tau_{\nu}} \, , 
\ee
where $T_S$ is the ``spin temperature" of the hydrogen atoms in the
gas cloud, and $T_{\gamma}$ is the CMB temperature,
$\tau_{\nu}$ is the dimensionless optical depth, and $\nu$ is the frequency. The
second term describes the absorption of the primordial
CMB photons, the first term yields the emission.

The spin temperature determines the excitation level
of the hydrogen atoms and is related to the gas temperature $T_K$
in the wake via a collision coefficient $x_c$ (see \cite{Furl} for
the exact equation). We are interested in the difference 
$\delta T_b$ between (\ref{basic1}) and the CMB temperature.
redshifted to the present time:
\be \label{basic3}
\delta T_b(\nu) \, = \, \frac{T_b(\nu)-T_{\gamma}(\nu) }{1 + z} \,  ,
\ee
which yields
\be
\delta T_b(\nu) \, = \, T_S \frac{x_c}{1 + x_c} \bigl( 1 - \frac{T_{\gamma}}{T_K} \bigr) 
\tau_{\nu} (1 + z)^{-1}  \, .
\ee
The optical depth is proportional to the width of the wake and to the line profile,
the broadening of the line due to the varying time of travel of photons emitted
at different points in the wake. The line profile is inverse proportional to the
width. Hence, the effect of the width (and hence of $G \mu$) cancels out of
the amplitude. Inserting numbers, one obtains
\be
\delta T_b(\nu) \, \sim  200 mK    \,\,\, {\rm for} \,\,\,,\, z + 1 = 30  \, ,
\ee
a very large effect (for details the reader is referred to \cite{Holder2}).

Since string wakes produce such a distinctive geometrical pattern, one
should search for strings making use of statistics which are sensitive
to this pattern. The use of Minkowski functionals for this purpose has
been explored in \cite{Evan}. One could also apply edge detection algorithms.

The calculations summarized above were done neglecting the thermal
velocities of the baryons before they start falling into the wake.
At early redshifts and for low values of $G \mu$ this is clearly not justified.
Including these initial thermal velocities will lead to the wake being
diffuse. This will increase the width of the wake (we speak of a
``diffuse string wake"), but will decrease the
pixel by pixel amplitude. The effect at a point in the sky integrated over
redshift remains the same, as was studied in \cite{Oscar2}. 

What range of values of $G \mu$ might be visible in 21cm surveys?
A conservative criterion for visibility is that the pixel signal due to
a string wake be larger than the expected pixel noise. The
study of \cite{Oscar2} indicates that values of $G \mu$ one
order of magnitude smaller than the current limit might be visible.
This is depicted in Figure 8 (taken from \cite{Oscar2})
which shows the string wave signal
as a function of $G \mu$ for various values in the redshift $z_i$
when the wake is laid down, and for a fixed reshift $z = 20$
when our past light cone intersects the wake, compared to the
noise (the almost horizontal lines to the top and bottom of the
horizontal axis). Note that the noise is a increasing function as the
pixel size decreases. Low values of $G \mu$ require smaller pixel
sizes and hence the noise level will be higher.  Making use of
algorithms sensitive to the particular geometrical pattern of the
string wake signal will allow us to probe significantly smaller
values of $G \mu$.

\begin{figure}
\includegraphics[height=5cm]{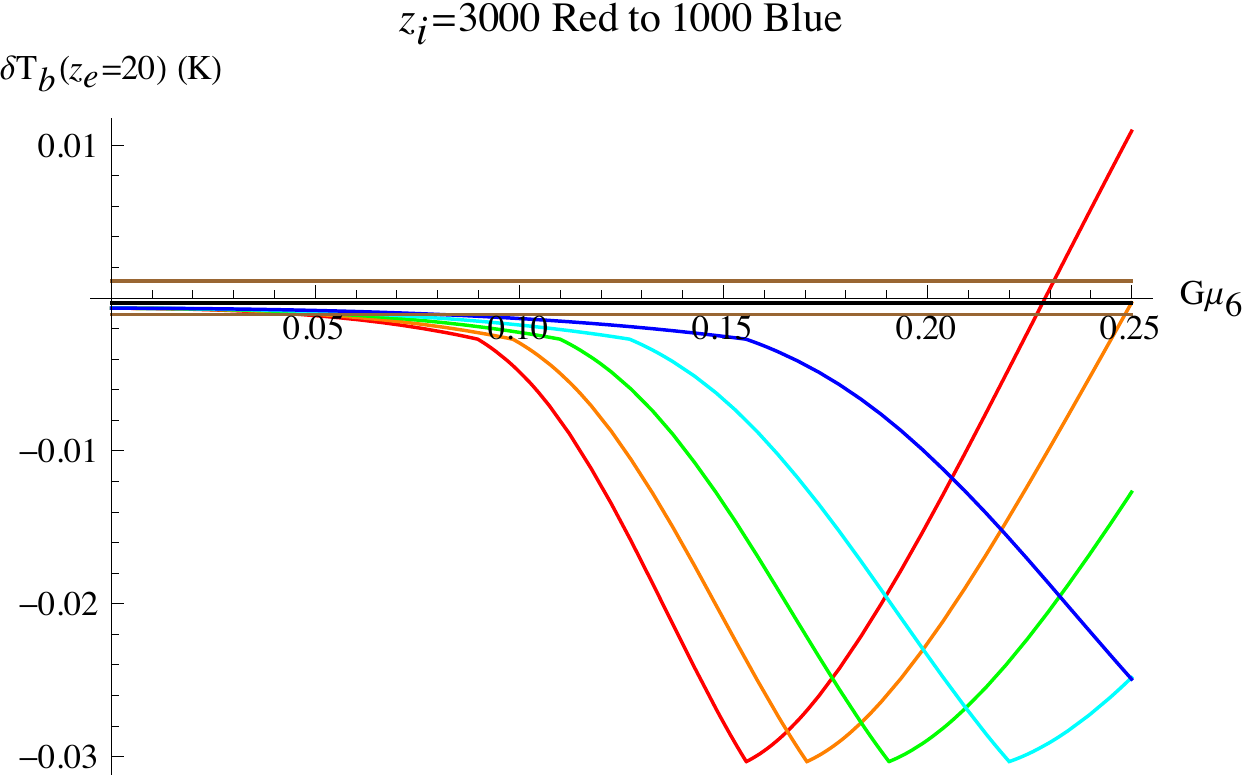}
\caption{Pixel amplitude of the string wake signal (vertical
axis) as a function of $G \mu$ (horizontal axis)
for various values of the wake formation redshift
$z_i$, for the past light cone crossing the wake at redshift $20$,
just before rionization. For very large values of $G \mu$ the kinetic 
temperature of the baryons in the wake is larger than the CMB
temperature and the string signal is in emission. For smaller values
of $G \mu$ the signal is in absorption. The string signal would
asymptote to a constant amplitude were it not for the effect of
diffusion. This leads to a decrease of the string wake signal for
very low values of the tension. As the figure shows, the string
signal remains larger than the noise (the almost horizontal lines)
to values of $G \mu$ one order of magnitude smaller than the current
upper bound.} 
\label{fig8}
\end{figure}

We close this section with a couple of final comments. First, the amplitude of the
21cm signal from a cosmic string loop is, as was studied in
\cite{Pagano} larger by a factor of 16
compared to the amplitude of the wake signal, the reason being
that the overdensity is larger by that factor (contraction occurs
in three instead of just one dimension). On the other hand, there
is no distincitive geometrical signature of the loop signal, and hence
it will be more difficult to tease apart the loop signal from point
source noise. Second, the two-dimensional angular power spectrum of the
string wake signal has also been calculated \cite{Oscar3}. 
However, the result does not show specific stringy features.
 
\section{Conclusions}

Since the seeds of the cosmological fluctuations which we observe today
have been laid down in the very early universe, these fluctuations may
carry imprints of the physics at the earliest times, and this means the
physics at highest energy scales. As an example, we have studied how
some observational windows can be used to probe for the possible
existence of cosmic strings produced by particle physics theories
beyond the Standard Model. 

One of the important conclusions is that the signals of strings are
more prominent in position space than in Fourier space. Hence, to
find or constrain strings it is important to use good position space
statistics (some attempts were described in the text). For CMB temperature
anisotropy maps the distinctive signal of string wakes is line discontinuities,
for CMB polarization maps it is rectangles on the sky with a polarization
amplitude which increases linearly in one direction, and whose polarization
angle is roughly constant. In particular, there is direct B-mode polarization.
In the case of 21cm redshift maps, string wakes lead to wedges which are
extended across the angular directions but thin along the redshift axis.
The amplitude of the position space signals is typically independent of
the unknown constant $N$ which describes the string scaling solution,
except when overlaps on the sky of the signals of different string wakes
are important.

We have assumed here, for simplicity,
that the strings are topologically stable and non-superconducting.
Two interesting avenues of extension are to consider superconducting
strings \cite{Witten2} for which electromagnetic interactions are important
and will lead to a rather different cosmological scenario (see e.g. \cite{SCSreview}
for a review) and non-topological strings.

Non-topological strings (see \cite{Ana} for a review) are string solutions
which are unstable in the vacuum. However, as first discussed in \cite{Nagasawa}
and more recently in \cite{Karouby} such strings can be stabilized in a
plasma. In fact, even the Standard Model of particle physics admits
strings (namely the electroweak Z-string and the pion string of
the low-energy sigma model of QCD) which can be stabilized
by the electromagnetic plasma present before the time of recombination. 
In this case, they will form in the early universe like topological
strings, and they can play a similar cosmological role.

Strings are not the only type of defect solutions. We have focused on 
string solutions since they are the cosmologically
most interesting ones. Models with domain walls are ruled out if the
energy scale is beyond the scale of the LHC since a single domain wall
would overclose the universe \cite{DWproblem}. Gauge monopoles are
also ruled out by overclosure considerations for high energy scale
walls \cite{Preskill}.

Finally, it is important to mention that all our studies of cosmic string
signals have been in the context of a one-scale toy model for the
distribution of strings. It would be interesting to study the string
signals using the inputs of actual string evolution simulations. There
are clearly lots of interesting projects to be undertaken.
 
\vskip0.5cm 

\centerline{\bf Acknowledgments}
 
I wish to thank Prof. Pauchy Hwang for the invitation to contribute to
one of the first issues of ``Universe". I wish to thank Rebecca Danos 
for permission to use various figures drawn from \cite{Danos, Holder1, Holder2},
and to Oscar Hernandez for permission to use Figure 8 which is taken from
\cite{Oscar2}. This research has been supported in part by an NSERC Discovery Grant
and by funds from the Canada Research Chair program.

\end{document}